\documentclass[twocolumn,epjc3]{svjour3}  

\journalname{Eur. Phys. J. A}
\usepackage[utf8]{inputenc}
\usepackage{epsfig,amsfonts,amssymb,latexsym,amsmath,bm,color,array,epsfig,graphics}

\usepackage{enumitem}
\usepackage{amsmath}
\usepackage{amssymb}
\usepackage{lastpage}
\usepackage{graphicx}
\usepackage{subfig}
\usepackage{hyperref}
\hypersetup{colorlinks=true,linkcolor=blue,citecolor=blue,menucolor=black,urlcolor=blue,filecolor=blue}

\newcommand{\be}{\begin{equation}}
\newcommand{\ee}{\end{equation}}
\newcommand{\bea}{\begin{eqnarray}}
\newcommand{\eea}{\end{eqnarray}}

\newcommand{\new}{\color{black}}
\newcommand{\newnew}{\color{black}}

\newcommand{\Ds}{D_{s0}^*(2317)}
\newcommand{\Zb}{Z_b(10610)}
\newcommand{\Zbs}{Z_b(10650)}

\usepackage{units} 
\usepackage{subfig} 

\usepackage{xargs}
\usepackage[usenames,dvipsnames]{xcolor}
\usepackage[colorinlistoftodos,prependcaption,textsize=scriptsize]{todonotes}
\newcommandx{\unsure}[2][1=]{\todo[linecolor=OliveGreen,backgroundcolor=OliveGreen!25,bordercolor=OliveGreen,#1]{#2}}
\newcommandx{\change}[2][1=]{\todo[linecolor=red,backgroundcolor=red!25,bordercolor=red,#1]{#2}}
\newcommandx{\info}[2][1=]{\todo[linecolor=blue,backgroundcolor=blue!25,bordercolor=blue,#1]{#2}}
\newcommandx{\improve}[2][1=]{\todo[linecolor=Plum,backgroundcolor=Plum!25,bordercolor=Plum,#1]{#2}}
\newcommandx{\notforall}[2][1=]{\todo[disable,#1]{#2}}

\smartqed 
\RequirePackage{mathptmx} 
\RequirePackage[numbers,sort&compress]{natbib}

\usepackage[left=2cm,right=2cm,top=3cm,bottom=3cm]{geometry}
\graphicspath{{Figures/}}

\begin{document}
\emergencystretch 3em

\title{On the nature of near-threshold bound and virtual states}

\author{ 
Inka Matuschek\thanksref{add1} \and 
Vadim Baru\thanksref{add2,add3,add4} \and 
Feng-Kun Guo\thanksref{add5,add6}   \and 
Christoph Hanhart\thanksref{add1} 
}       

\institute{Forschungszentrum J\"ulich, Institute for Advanced Simulation, Institut f\"ur Kernphysik and
J\"ulich Center for Hadron Physics, D-52425 J\"ulich, Germany \label{add1}
\and
Helmholtz-Institut f\"ur Strahlen- und Kernphysik and Bethe Center for Theoretical Physics, Universit\"at Bonn, D-53115 Bonn, Germany \label{add2}
\and
Institute for Theoretical and Experimental Physics NRC (Kurchatov Institute), Moscow 117218, Russia \label{add3}
\and
P.N. Lebedev Physical Institute of the Russian Academy of Sciences, 119991, Leninskiy Prospect 53, Moscow, Russia \label{add4}
\and
CAS Key Laboratory of Theoretical Physics, Institute of Theoretical Physics, Chinese Academy of Sciences,  Beijing 100190, China   \label{add5}
\and
School of Physical Sciences, University of Chinese Academy of Sciences, Beijing 100049, China\label{add6}
}

\date{}
\maketitle

\abstract{
Physical states are characterised uniquely by their pole positions and the corresponding residues.
Accordingly, in those parameters also the nature of the states should be encoded. For bound states
(poles on the real $s$-axis below the lowest threshold on the physical sheet) there is an established 
criterion formulated originally by Weinberg in the
1960s, which allows one to estimate the amount of compact and molecular components in a given state.  We demonstrate in this paper that this criterion can be straightforwardly extended
to shallow virtual states (poles on the real $s$-axis below the lowest threshold on the unphysical sheet) which should be
 classified as molecular.
We argue that predominantly non-molecular or compact states 
exist either as bound states or as resonances (poles on the unphysical sheet off the real energy axis) but not as virtual states. 
We also discuss the limitations of the mentioned classification scheme.
} 

\section{Introduction}
\label{intro}

The observations of the $D_{s0}^*(2317)$~\cite{Aubert:2003fg}, which is slightly below the $DK$ threshold, and the charmonium-like meson $X(3872)$ in 2003~\cite{Choi:2003ue}, which lies remarkably close to the $D^0\bar{D}^{*0}$ threshold, 
sparked a renewed interest in heavy meson spectroscopy, since their properties are in conflict  with the predictions of the quark model. 
Since then a large number of exotic hadrons which cannot be accommodated by conventional quark models, 
were discovered and are still being found in such facilities as BaBar, Belle, BESIII and LHCb. Those in the heavy quarkonium mass region are called $XYZ$ states.
Various models were developed for describing these exotic states. The proposals include hadro-quarkonia, hybrids, tetraquarks, molecular states and kinematical effects,  for recent reviews see, for example, Refs.~\cite{Lebed:2016hpi,Esposito:2016noz,Ali:2017jda,Guo:2017jvc,Olsen:2017bmm,Liu:2019zoy,Brambilla:2019esw,Guo:2019twa}. All these states have in common, that their masses are located above the first  heavy-quark 
open-flavor threshold, $\bar{D}D$ or $\bar{B}B$, respectively. Moreover, many masses lie close to some open-flavor threshold, 
which makes these states natural candidates for hadronic molecules. 

Unrevealing the nature of the $XYZ$ states is of high interest for it will shed light on how quantum chromodynamics (QCD) forms hadrons.
To form the basis for this quest, one needs to provide  theoretically sound definitions of the different structures. In the 1960's Weinberg found a way to discriminate between composite (or molecular)  and elementary (or compact)  near-threshold bound states in the weak-binding limit~\cite{Weinberg:1962hj,Weinberg:1963zza,Weinberg:1965zz}.
In particular,  he showed that the compositeness is given by $1-Z$, where $Z$ is the field renormalization constant of a state. 
Moreover, he derived that the  residue at the pole in the  scattering amplitude of two hadrons, which can couple to this state, directly measures $Z$ and obtained relations between $Z$, the scattering length $a$ and the effective range $r$. 
Then, he applied this model-independent scheme to the deuteron and demonstrated that the deuteron indeed is not an elementary particle.

{\new Before we proceed, it is useful to clarify the nomenclature and the physical meaning of certain near threshold
pole locations. Physical  asymptotic states, in the context of scattering theory often referred to as bound states, are stable. They can be used as beams or targets  in experimental setups and can also be measured
as final states in experiments. In the scattering amplitude, such states show up as poles on the physical Riemann sheet below the lowest
threshold in the complex energy or $s$-plane.
Into this class falls for instance the deuteron. Very similar features have very narrow (which means long-lived) unstable states,
if their corresponding pole sits on the first sheet with respect to the nearby threshold and the inelastic threshold is 
remote~\cite{Baru:2003qq}. Here, one often speaks of quasi-bound states.
To be distinguished from (quasi-)bound states are poles on the unphysical sheet(s). There are on the one hand virtual
states (located on the real axis of the unphysical sheet of the energy plane below threshold) and on the other hand resonances
(located in the complex plane of the unphysical sheet  with a nonvanishing imaginary part). Resonance poles need to appear in pairs thanks to the Schwarz reflection
principle. Clearly, neither of these states appear as asymptotic states and the associated wave functions are not normalisable, but they can have 
a significant impact on observables. }
Especially because most of the states of interest are indeed resonances, quasi-bound or virtual states and {\newnew the distinction between them without a careful analysis can be subtle (for a detailed discussion see, e.g.,  Ref.~\cite{Baru:2003qq,Kamiya:2016oao}),   it appears necessary to derive model-independent tools applicable to access the nature of a state for all these scenarios.  }

The mentioned Weinberg-compositeness criterion relies on the normalization of the wave function. Accordingly, 
it is formally applicable only to bound states. In recent years, however,  much research has been devoted to a possible generalization of this approach,
e.g. the generalization to coupled channels was done in Ref.~\cite{Baru:2003qq}. Of special interest is a consistent definition of the compositeness for states corresponding to poles on the unphysical sheet, i.e., resonances and virtual states. 
Related studies can be found in Refs.~\cite{Gamermann:2009uq,Sekihara:2016xnq,Sekihara:2015gvw,Kamiya:2015aea,Kamiya:2016oao,Aceti:2012dd,Aceti:2014ala,Guo:2016wpy,Guo:2015daa,Hyodo:2013nka,Hyodo:2011qc,Kang:2016ezb,Oller:2017alp,Sekihara:2014kya,Xiao:2016wbs,Xiao:2016dsx,Hyodo:2013iga}. 
Most of these works, however, focus on  resonances and cannot be applied to  virtual states.

It is a general understanding that virtual states are essentially molecular, although to the best of our knowledge so far no proper proof of this assertion was provided.
In this paper, we provide this missing proof. For this, we use the close connection between bound and virtual states in two-body $S$-wave scattering to develop a reliable criterion 
to estimate the compositeness of near-threshold virtual states. 
The paper is organized as follows. In Sec.~\ref{Sec:Z}, we
revisit the Weinberg criterion and discuss possible extensions, and in Sec.~\ref{Sec:poles} we employ the relation between the pole positions  and  the compositeness to demonstrate that virtual states can be considered mostly molecular. 
Then in Sec. \ref{Sec:appl}, we discuss applications.    The first state we study is the shallow virtual state that appears in the $^1S_0$-channel in  nucleon-nucleon scattering. 
We also analyze lattice results for NN scattering at the unphysical pion mass $m_{\pi} =450$~MeV. 
Then, we proceed to the $DK$ system focusing in particular on the $\Ds$ pole trajectory when some fundamental constant such as the quark mass is varied.
Employing  the results of Ref.~\cite{Liu:2012zya}, we discuss the transition of a bound state in this  $DK$ channel to a virtual state when the strange-quark mass (or simply the kaon mass) is varied.
Finally, we comment on the exotic $\Zb$ and $\Zbs$ states which allows us to also discuss the limitations of our approach.
We close with a short summary and outlook.

\section{Weinberg Criterion and Extensions}
\label{Sec:Z}

\subsection{Field renormalization factor for bound states}
\label{Zbound}

We start with the derivation of the Weinberg compositeness criterion  following Ref.~\cite{Baru:2003qq},  see also Ref.~\cite{Guo:2017jvc} for a review.
Since we focus on near-threshold states, a nonrelativistic formalism is justified.
Moreover, to simplify the notation, the argument is presented for scalar particles only.
Then, a physical bound state, realized as a pole on the first (physical) sheet, is assumed to consist of a bare compact state $| \psi_0 \rangle$ and a two-hadron channel $| h_1 h_2 \rangle_{\vec{p}}$,
characterized by the relative momentum $\vec{p}$ of the two hadrons:
\begin{equation}
	| \Psi \rangle
		= \left(
			\begin{matrix}
				\lambda \: | \psi_0 \rangle \\
				\chi(\vec{p}) \: | h_1 h_2 \rangle_{\vec{p}}
			\end{matrix}
		  \right).
\end{equation}
The probability of finding the bare state in the physical state is given by
\begin{equation}
	\left| \langle \psi_0 | \Psi \rangle \right|^2 = \lambda^2.
\end{equation}
This is the quantity we are interested in, since $(1-\lambda^2)$ gives the compositeness of the physical state. The interaction is governed by the Hamiltonian
\begin{equation}
	\hat{\textbf{H}}
		= \left(
			\begin{matrix}
				\hat{H}_c & \hat{V} \\
				\hat{V} & \hat{H}_{hh}^0
			\end{matrix}
		  \right).
\end{equation}
We assume that a proper field redefinition has been performed, such that the $h_1h_2$ interaction was eliminated from the Lagrangian. 
Therefore, $\hat{H}_{hh}^0$ contains only the two-hadron kinetic term
$\vec p^2/(2\mu)$, where $\mu=m_1 m_2/(m_1+m_2)$ denotes the reduced
mass of $h_1$ and $h_2$. Such kind of field transformation is always
possible at least if there is only  a single pole on the first
sheet~\cite{Weinberg:1962hj}, see also Ref.~\cite{Baru:2010ww}, in which the derivation of the compositeness criterion was done including the interaction in the hadronic channel explicitly.
Employing the Schr\"odinger equation, $\hat{\textbf{H}} | \Psi \rangle = E | \Psi \rangle$, one finds
\begin{align}
	E \chi(\vec{p}) 
		&= \: _{\vec{p}}\langle h_1 h_2 | \hat{\textbf{H}} | \Psi \rangle \nonumber \\
		&= \: _{\vec{p}}\langle h_1 h_2 | \hat{V} \lambda | \psi_0 \rangle \nonumber\\
		& \quad
			+ \int \!\! \frac{d^3p^\prime}{(2\pi)^3} {}  _{\vec{p}}\langle h_1 h_2 | \hat{H}_{hh}^0 \chi(\vec{p}^{\prime}) | h_1 h_2 \rangle_{\vec{p}^{\prime}}  \nonumber\\
		&= \lambda f(p^2) + \frac{p^2}{2\mu} \chi(\vec{p}),
\end{align}
where we introduced the transition form factor $f(p^2) = \langle \psi_0 | \hat{V} | h_1 h_2 \rangle_{\vec{p}}$. This yields
\begin{equation}
	\chi(p) = \lambda \frac{f(p^2)}{E-{p^2}/{(2\mu)}}.
\end{equation}
Employing this relation the normalization condition of the physical state can be expressed as
\begin{equation}
	1 
		= \langle \Psi | \Psi \rangle
		= \lambda^2 \left( 1 + \int  \!\! \frac{d^3p}{(2\pi)^3} \frac{f^2(p^2)}{(E_B + {p^2}/{(2\mu)})^2} \right) \ ,
\label{eq:Weinberg_NormalizationPsi}
\end{equation}
where the binding energy is defined via $E_B=m_1+m_2-M$, with $M$ for the mass of the state under
investigation.
We can compare this to the definition of the field renormalization $Z$ in the nonrelativistic theory
\be\label{GE}
	G(E) = \frac{1}{E-E_0-\Sigma(E)} = \frac{Z}{E+E_B}+ {\cal O}\left((E+E_B)^2\right),
\ee
where
\be
	 -E_B{=}E_0+\Sigma(-E_B), \quad  \ Z {=} \frac{1}{1-\left.\partial\Sigma/\partial E\right|_{E=-E_B}}. 
	 \label{eq:Zselfenergy}
 \ee
Since
\begin{equation}
	\Sigma(E) = \int  \!\! \frac{d^3p}{(2\pi)^3} \frac{f^2(p^2)}{E-{p^2}/{(2\mu)+i 0}} ,
\label{eq: Selfenergy}
\end{equation}
it follows that
\begin{equation}
	Z = \left(1 + \int  \!\! \frac{d^3p}{(2\pi)^3} \frac{f^2(p^2)}{(E_B+{p^2}/{(2\mu)})^2}\right)^{-1} \ .
	\label{eq:Zdef}
\end{equation}
Comparing Eq.~(\ref{eq:Zdef}) to Eq.~(\ref{eq:Weinberg_NormalizationPsi}) yields
\begin{equation}
	Z = \lambda^2 \ .
	\label{eq:Z=xi2}
\end{equation}
The compositeness is therefore given by $1-Z$.

It is well known from  textbooks on  quantum field theory that the renormalization field factor is  a scheme and even regularization dependent quantity. 
Indeed, those pieces of $Z$ that are analytic in $E$ are scheme dependent and need to be fixed by some
renormalization condition. However,  here we are mostly interested in the weak-binding limit defined via $\gamma \ll \beta$, where
$\gamma=\sqrt{2\mu E_B}$ denotes the binding momentum and $\beta$ the closest non-analyticity of the system not 
related to the threshold under investigation. Often $1/\beta$ can be identified with the range of forces or with the leading left-hand cut of the potential but, as will be discussed
below, it can also refer to the distance to the next higher threshold.
In this limit  for $S$-waves, a non-analytic model-independent piece dominates $Z$.
To see this, one may observe that $\beta$ sets the scale for the momentum variation of $f(p^2)$ such that we may write
\begin{align*}
	\int \frac{f^2(p^2)d^3p}{(E_B+{p^2}/{(2\mu)})^2} &=
	 16 \mu^2 \pi
			\int_0^\infty \!\!\!\!\!\!   \frac{f^2(p^2)p^2 dp}{(p+i\gamma)^2 (p-i\gamma)^2}  \\
		&= 4 \mu^2 \pi^2 \frac{f^2(-\gamma^2)}{\gamma} \left[1 + \mathcal{O}\left(\frac{\gamma}{\beta}\right)\right] \ .
\end{align*}
The terms that result from the singularities of the transition form factor are contained in the last term in the last line, which
should be small in the weak binding limit. The scheme and scale dependent part is contained in the suppressed $\mathcal{O}(\gamma/\beta)$ terms. Using $f(-\gamma^2)=g_0^2$ for the unrenormalized 
coupling constant, we get from Eq.~(\ref{eq:Zdef}) 
\begin{equation}
	\frac{1}{Z} - 1 = \frac{\mu^2 g_0^2}{2\pi\gamma} + \mathcal{O}\left(\frac{\gamma}{\beta}\right).
\end{equation}
Hence, in the weak binding limit ($\gamma \ll \beta$), the quantity $Z$, which is a measure of the compositeness, is dominated by a calculable, non-analytic piece. The effective coupling 
\begin{equation}
	g^2=Zg_0^2 = \frac{2\pi\gamma}{\mu^2}(1-Z)
	\label{eq:g}
\end{equation}
is the residue of the scattering amplitude and is therefore a measurable quantity.

Note that this analysis is model independent only if a  state couples with  a two-body continuum channel in an  $S$-wave.  Indeed, since the centrifugal barrier $p^{2L}$ needs to be included in the integral in Eq.~(\ref{eq:Zdef}),  for higher partial waves, already the leading piece of the integral  depends
on the  model-dependent parameter $\beta$. 
In addition, it is imperative that the constituents of the bound state are narrow, since otherwise the bound state would be broad as well~\cite{Filin:2010se}. The narrowness is characterized by $\Gamma$ with $\Gamma \ll \beta$, which is understandable as otherwise the constituents would not have enough time to interact to form a bound state~\cite{Guo:2011dd}.

With the notation introduced above, the scattering matrix with a pole on the physical sheet at $E=-E_B$ reads 
\begin{equation}
 T(E) = \frac{g_0^2}{E+E_B+\frac{g_0^2\mu}{2\pi}(i k +\gamma)}  + \mbox{non-pole terms},
 \label{Tg0}
\end{equation}
where we used that  Im$(\Sigma(E))=-i\mu k g_0^2/(2\pi)$, with $E=k^2/(2\mu)$.
To relate $g_0$ and  $Z$ to observables, we match Eq.~(\ref{Tg0}) to the effective range expansion (ERE) defined as
\begin{equation}
 T(E)=-\frac{2\pi}{\mu}\frac{1}{1/a + (r/2)k^2-i k} \ ,
 \label{eq:T_ERE}
\end{equation}
where $a$ ($r$) denotes the scattering length (the effective range). This yields
\begin{align}
	a &= -2 \: \frac{1-Z}{2-Z} \frac1\gamma + \mathcal{O}(1/\beta), \nonumber \\
	r &= - \frac{Z}{1-Z}  \frac1\gamma + \mathcal{O}(1/\beta) \: .
\label{eq: ERE Parameters from Z}
\end{align}
To see how these equations work, consider the two extreme cases of a pure molecule and a purely compact state. 
The former implies $Z=0$ and thus the absolute value of the scattering length gets maximal, $a = -1/\gamma$, while $r = \mathcal{O}(1/\beta)$ is of natural size and typically positive, although below we will discuss an example of a predominantly
mole\-cular state with a negative effective range. On the other hand, in case of a purely compact bound state the scattering length takes a natural value,
 $a = -\mathcal{O}(1/\beta)$, and the effective range gets unnaturally large and negative.
Solving Eqs.~(\ref{eq: ERE Parameters from Z}) for $Z$, in the zero-range approximation (neglecting the
$\mathcal{O}(1/\beta)$ terms),  one finds 
\begin{equation}
	1-Z = \sqrt{\frac{a}{a+2r}}=:X \ ,
\label{eq: Z from ERE}
\end{equation}
where we introduced the compositeness $X$. It follows directly from Eq.~(\ref{eq: ERE Parameters from Z})
that Eq.~(\ref{eq: Z from ERE}) holds only  when both $a$ and $r$ are negative. While the
former condition is correct as soon as the relevant pole is on the physical sheet, the latter signals
that in the derivation range corrections were neglected.

\begin{figure}[t]
\centering
\includegraphics[width=.2\textwidth]{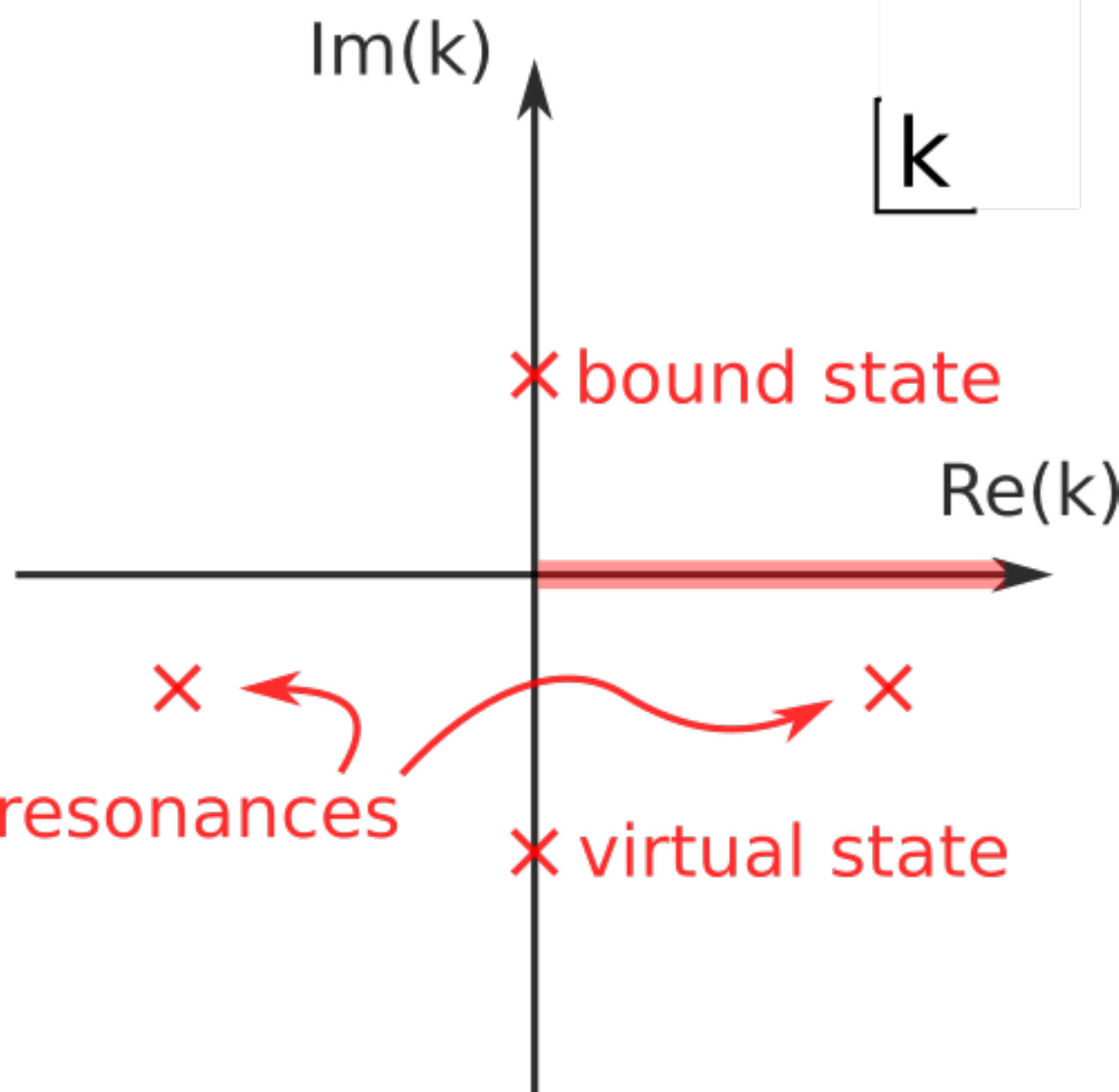}\\
\caption{Naming convention for the poles in the $k$-plane. The thick red line for
positive real valued $k$ marks the physical momenta in the scattering regime.}
\label{fig:PolesKPlaneNaming}
\end{figure}

\subsection{Possible extensions of compositeness beyond bound states}
\label{Sec:extension}

\begin{figure}
\centering
\includegraphics[width=.3\textwidth]{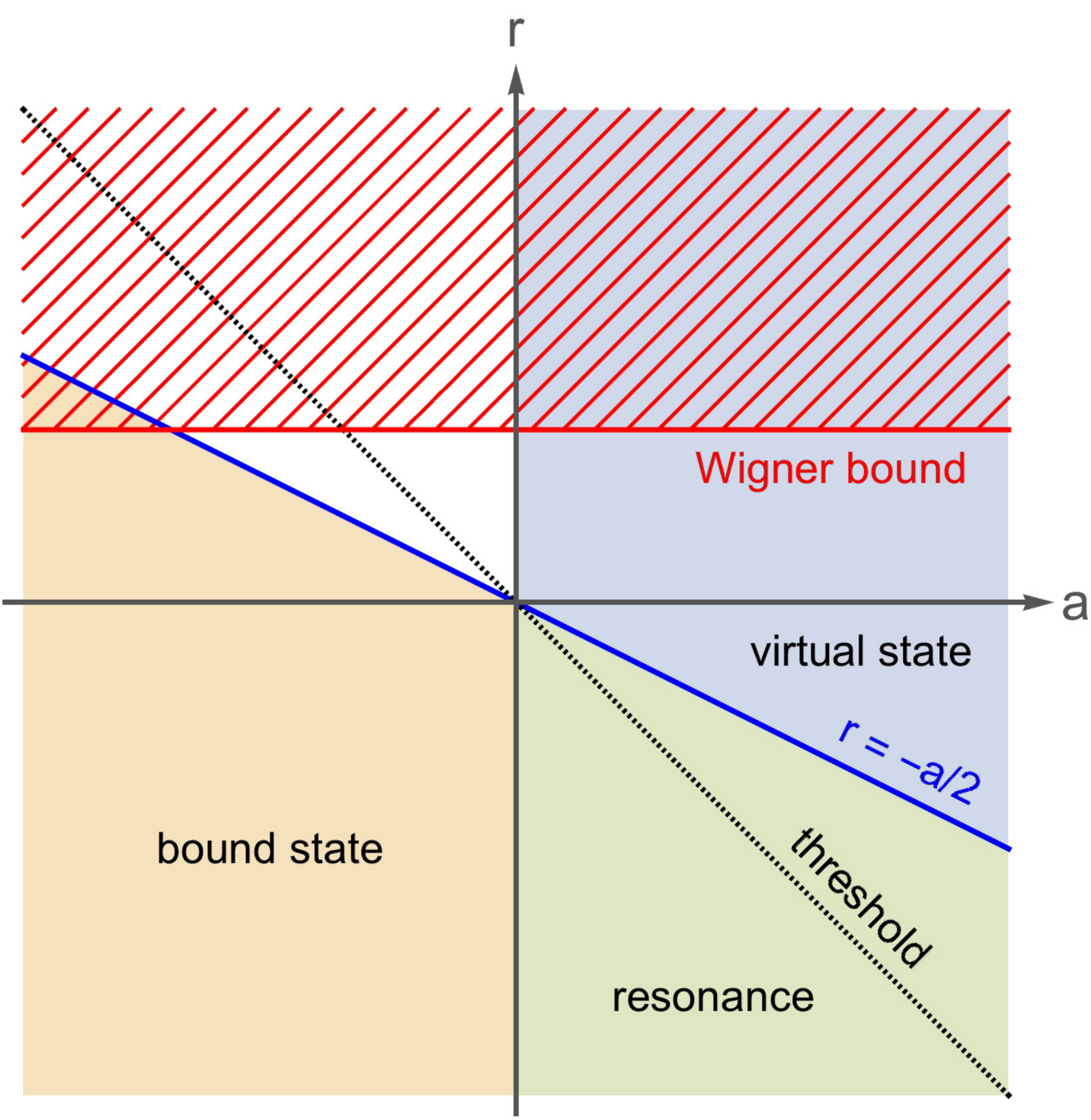}
\caption{Types of poles in the $r$--$a$ plane. The dotted line, located at $r=-a$, refers
to those poles that have a vanishing real part in the $E$ plane and
accordingly are located right at threshold. }
\label{fig:arpolelocations}
\end{figure}

Physical states are associated with poles of the $T$-matrix. From Eq.~(\ref{eq:T_ERE}) it follows that there are two poles located at
\begin{equation}
	k = \frac{i}{r} \left( 1 \pm \sqrt{1+\frac{2r}{a}} .\right)
\end{equation}
The leading pole is the pole closest to the physical axis. 
States with their leading pole on the positive imaginary momentum axis are called bound states, since by definition the sheet with
momenta that have positive imaginary parts refers to the physical sheet. On the other hand,
states on the negative imaginary momentum axis are called virtual states and all other physically allowed states resonances (see Fig.~\ref{fig:PolesKPlaneNaming}). 
Analyticity demands that resonance poles always appear in pairs, as shown in this figure. 
All of these possibilities can be related to different values  of the effective range parameters, as indicated in Fig.~\ref{fig:arpolelocations}.
Note that resonance poles are located above the corresponding threshold when studied in the
(energy)  $k^2$ plane (or the Mandelstam $s$ plane) only if the following conditions are fulfilled simultaneously: $|a|<|r|$, $r<0$ and $a>0$. The line $a=-r$ is also shown as the dotted line in Fig.~\ref{fig:arpolelocations}. 
The region above the $(r=-a/2)$-line for $a<0$ in  Fig.~\ref{fig:arpolelocations} is not carrying any name and is left white, for it refers to poles in the complex plane of the physical sheet. Such poles are unphysical since the resulting states would be at odds with causality.
Moreover, positive effective ranges that vastly exceed the range of forces also lead to a violation of causality --- this fact, represented in the figure as the red hatched area, is known as Wigner bound~\cite{Wigner:1955zz} (for a modern discussion of the subject see Ref.~\cite{Hammer:2009zh}), see \ref{sec:app}.

\begin{figure}
\centering
\parbox{6cm}{(a)

\includegraphics[width=.35\textwidth]{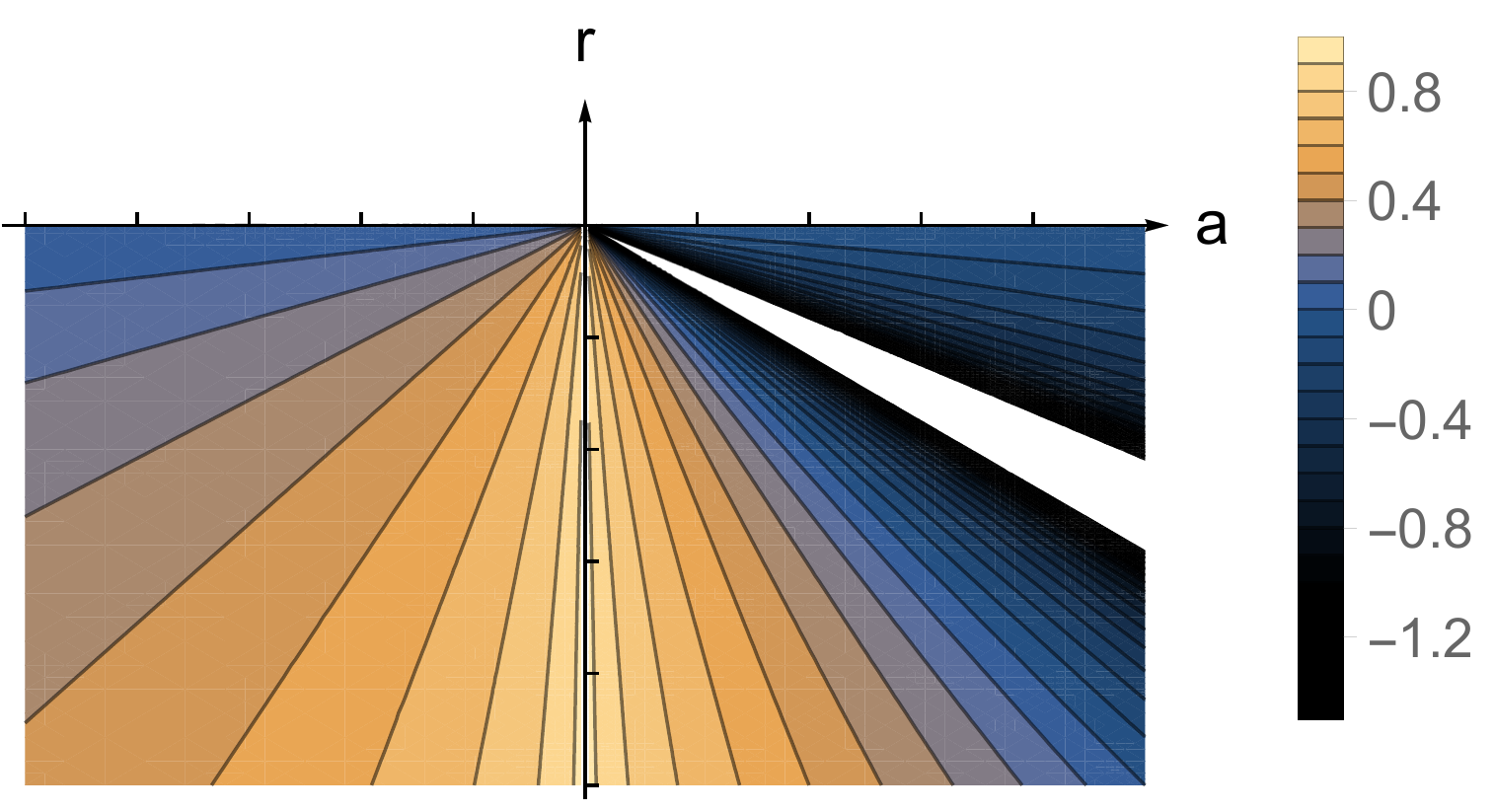}}

\vspace{0.5cm}

\centering
\parbox{6cm}{(b)

\includegraphics[width=.35\textwidth]{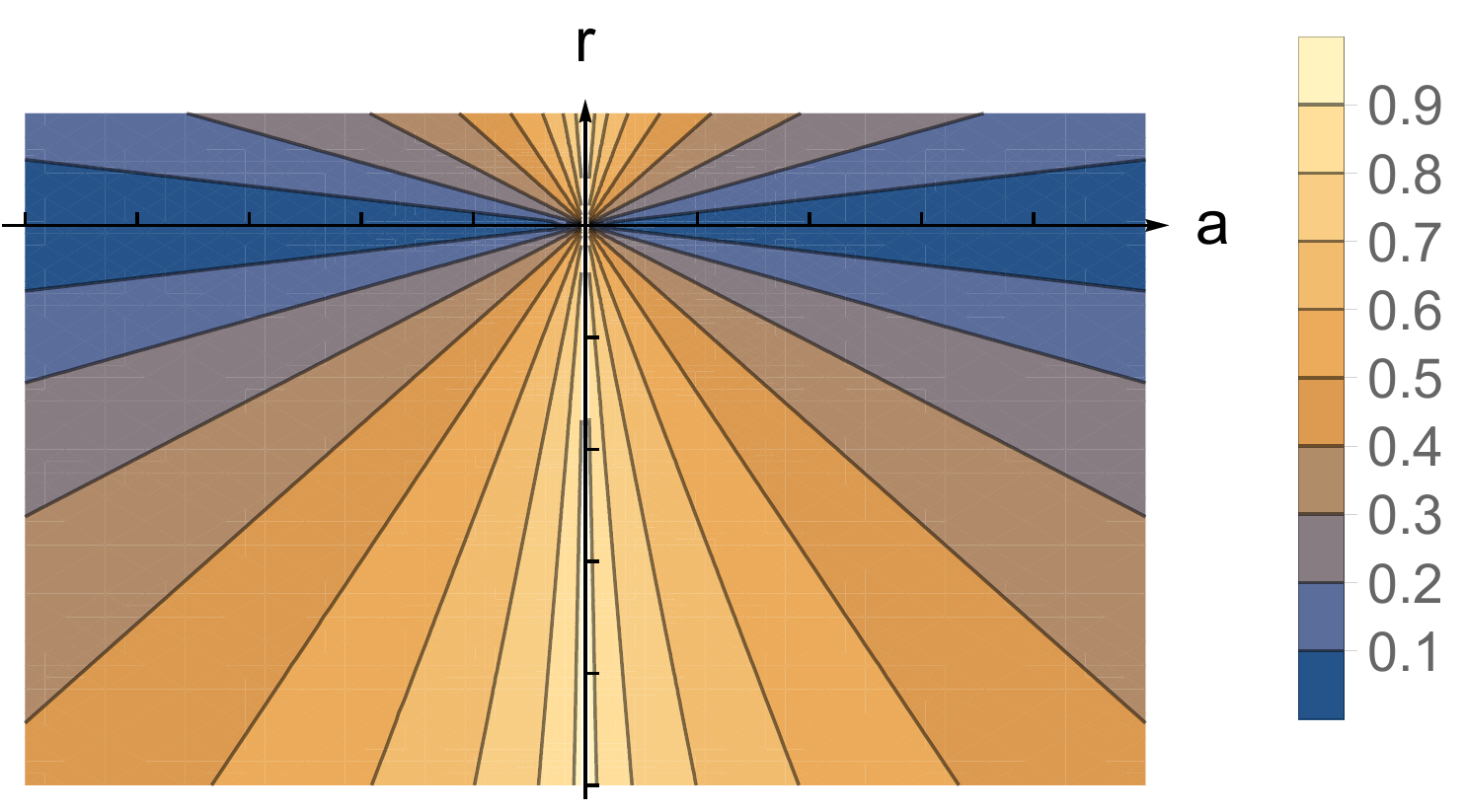}}

\caption{(a): $\bar{Z}_H$ as suggested in Ref.~\cite{Hyodo:2013iga}, with the regions where
the expression diverges left white; (b): $\bar{Z}_A$ as in Eq.~(\ref{eq: Zbar via a,r; own}).
In the lower left quadrant ($r<0, a<0$) the two prescriptions agree by construction.}
\label{fig:Z_via_X}
\end{figure}

\begin{figure}[b]
\centering
\includegraphics[width=.35\textwidth]{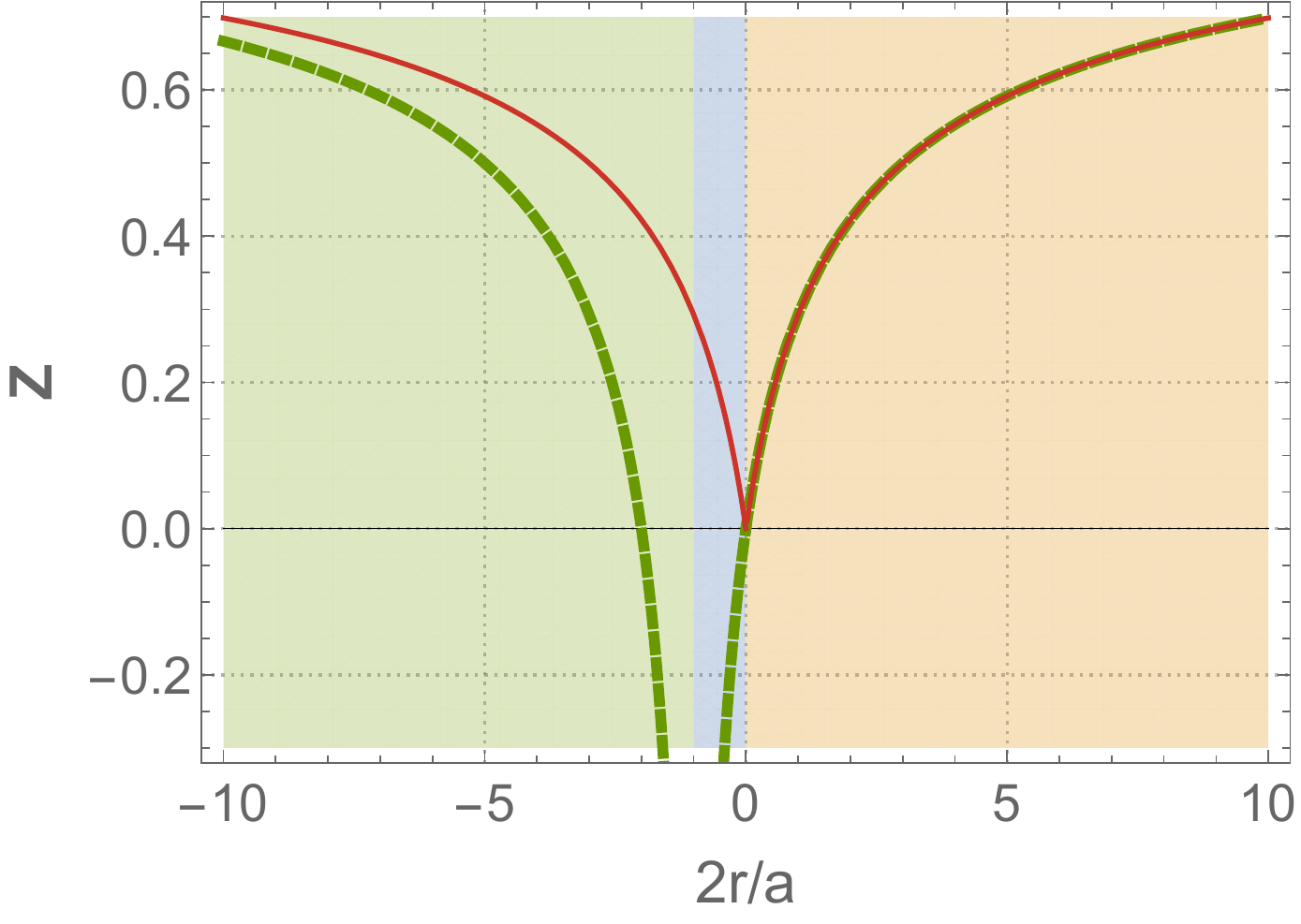}
\caption{$\bar{Z}_H$ (green dashed line) and $\bar{Z}_A$  (red solid line) versus $2r/a$. The horizontal thin (black) line corresponds to $Z=0$. When $2r/a > 0$ and $r$ is negative both prescriptions agree with each other and with the Weinberg's $Z$ from Eq.~\eqref{eq: Z from ERE}. The compactness $\bar{Z}_H$ has a pole at $2r/a=-1$. }
\label{fig:ZHvsZA}
\end{figure}

It is not trivial to extend the notion of compositeness to states other than bound states, since wave functions derived from poles on the unphysical sheet are not normalizable and the probabilistic interpretation is lost.  
Nevertheless, relying formally on  the definition of the field renormalization $Z$ in the nonrelativistic theory (Eqs.~\eqref{GE}-\eqref{eq:Zdef}), relations between $a,r,g^2$ and $Z$ can be derived also for a virtual state with a pole at $k = -i\gamma = -i\sqrt{2\mu E_B}$ and are similar to those of a bound state (given above in Eqs.~\eqref{eq:g} and \eqref{eq: ERE Parameters from Z}):
one simply replaces $\gamma$ in these equations by $-\gamma$ to get the relations for the virtual state with $Z$ given in Eqs.~\eqref{eq:Zselfenergy} and \eqref{eq: Z from ERE}.

Motivated by the fact that the absolute value of $X$ is confined within the interval $0 < |X| < 1$ also for resonances above threshold, a possible prescription for compositeness was provided in Ref.~\cite{Hyodo:2013iga}. 
Namely, it was suggested to define compositeness via a continuation of Eq.~(\ref{eq: Z from ERE}) as
\begin{equation}
	\bar{X}_H = 1-\bar Z_H = \sqrt{\left|\frac{a}{a+{2r}}\right|}  \ ,
	\label{eq:Hyodo}
\end{equation}
where the subindex $H$ is put to indicate that this continuation is from Ref.~\cite{Hyodo:2013iga} by T. Hyodo,  see also Ref.~\cite{Kang:2016ezb} for a related study. 
Since no range corrections are included, the range of applicability of Eq.~(\ref{eq:Hyodo}) is supposed to be restricted by $r < 0$. 
The behavior of this continuation in the $(a-r)$-plane is illustrated in Fig.~\ref{fig:Z_via_X}(a). By construction, for bound states ($a<0, r<0$) the quantity $\bar{X}_H$ agrees with the Weinberg's $X$ from Eq.~\eqref{eq: Z from ERE}.
However, as it is seen from Fig.~\ref{fig:Z_via_X}\,(a),  when $|r|  \to a$ for $r<0$ and $a>0$,    $\bar{X}_H \to 1$  and one would then conclude that the state in this case is purely molecular, even though we are far from the typical molecular regime corresponding to  $|a| \gg |r|$. 
Moreover,  in the range $a/2 \le |r| < a$, $\bar{X}_H$  gets larger than one and even diverges when $|r| $ approaches $a/2$.
Actually, this peculiar behaviour of $\bar{X}_H$ is a direct consequence of the pole in Eq.~\eqref{eq:Hyodo}  when $r= -a/2$, see also Figs.~\ref{fig:Z_via_X}\,(a) and \ref{fig:ZHvsZA}.

As the choice of continuation of the $X$ is  not unique, as an alternative to Eq.~\eqref{eq:Hyodo},  we introduce the quantity 
\begin{equation}
        \bar{X}_A = 1- \bar{Z}_A= \sqrt{\frac{1}{1 + \left| {2r}/{a} \right| }},
\label{eq: Zbar via a,r; own}
\end{equation}
where $\bar{Z}_A$ is illustrated in figures \ref{fig:Z_via_X}\,(b) and  \ref{fig:ZHvsZA}. While by construction both formulations agree in the regime where the Weinberg formula, Eq.~(\ref{eq: Z from ERE}), is applicable, they differ significantly elsewhere. 
Moreover, Eq.~(\ref{eq: Zbar via a,r; own}) is applicable in the full $(a-r)$-plane, while Eq.~(\ref{eq:Hyodo}) is not. 
Indeed, the quantity $\bar X_A$ is always normalized as $\bar X_A \in [0, 1]$    and  provides a smooth transition from the regime of molecules to the regime of compact states avoiding artificial poles. 
Therefore, this quantity can be used to estimate the compositeness for bound states, resonances and virtual states, see next section for a detailed discussion. 
While the  original $X$ from Eq.~\eqref{eq: Z from ERE}  was derived in the zero-range approximation only, that is for $r<0$ (see \ref{sec:app}),  $\bar X_A$ provides also reasonable results for $r>0$. 
Consider, for example, the case of the deuteron which is known as the archetypical example of molecular states~\cite{Weinberg:1965zz}.   
Indeed, the experimental values for the scattering length and the effective range  ($a = -5.41$~fm and $r = +1.75$~fm) are in agreement with the condition $|a| \gg |r|$ corresponding to predominantly 
composite objects.  Meanwhile, an attempt to use these parameters to estimate  $X$ naively, that is  from Eq.~\eqref{eq: Z from ERE}, fails badly yielding $Z\approx -0.7$ and $X\approx 1.7$.
This seems to be in contradiction with the expectations that the range corrections to $X$ should  scale as $\mathcal{O}(\gamma/\beta)$, which translates to about 30\% for the deuteron.  However, 
while the range corrections in the deuteron are indeed of natural size, they change the sign of the effective range and therefore Eq.~\eqref{eq: Z from ERE} is not applicable anymore.   On the other hand, using Eq.~\eqref{eq: Zbar via a,r; own}  gives  $\bar X_A= 0.8$, which includes the natural effect from the range corrections and is completely consistent with our expectations for this molecular state.
Note, however,  that $\bar X_A$ should still be understood as the leading-order approximation to the compositeness, since range corrections cannot be included in $\bar X_A$  in a model-independent way.

In what follows, we discuss pole trajectories focusing on the limiting cases of  the transitions from bound states to virtual states and to resonances. 
We will argue, that as long as range corrections can be neglected, shallow virtual states are indeed dominated by their two-hadron component, while narrow near-threshold resonances are of predominantly compact nature.

\section{Pole Trajectories and Compositeness of States}
\label{Sec:poles}

In order to study the origin of a state, it is instructive to follow the trajectories of its poles as they move from the physical sheet to the unphysical sheet when some QCD parameter (such as a quark mass) is varied.  
Before studying this on a concrete example, we discuss some general features of the trajectories under the assumption that the largest impact on the poles  is provided by a change in the scattering length,  while the  effective range remains constant. This kind of study is motivated by the trajectories of the $f_0(500)$ reported in Ref.~\cite{Hanhart:2014ssa}.  
Also in $NN$ scattering, the analysis of lattice data from Refs.~\cite{Baru:2015ira,Baru:2016evv}  shows that the effective ranges  in the $^1S_0$ and $^3S_1$ channels vary with the pion mass  much more smoothly than the  corresponding scattering lengths.

Let us first look at a system with a shallow bound state controlled by  the condition  $ |{r}| \ll |{a}|$, where  $a$ is large and negative.
The corresponding state is predominantly molecular.  Then, to leading order (LO) in $ |{r}/ {a}|\ll 1$, the scattering amplitude in the effective range approximation possesses two well separated poles
\be\label{Eq:poles}
k_1= -\frac{i}{a} \left[1+ \mathcal{O}\left(\frac{r}{a}\right) \right] , \quad  k_2= \frac{2i}{r}\left[1+ \mathcal{O}\left(\frac{r}{a}\right) \right].  
\ee
The  bound state pole $k_1$ resides close to the threshold and, therefore, leaves a remarkable imprint in the observables. Being driven by the scattering length, this pole is  stable  against the inclusion  of the range corrections $\sim  \mathcal{O}(1/\beta)$  in the hadronic potential, which can be systematically included within some  low-energy expansion.  
On the contrary,  the pole $k_2$  strongly depends on the model  used --- its actual location will change significantly if the range  corrections  as well as  higher-order terms in the effective range expansion are taken into account, since $r\sim 1/\beta$.  However, as long as  $| {r} |\ll |{a}|$,  the pole $k_2$ is always remote and therefore has at most a very small effect on observables.

Employing Eqs.~\eqref{eq: ERE Parameters from Z}, neglecting terms of $\mathcal{O}(1/\beta)$, the relevant pole  $k_1$   can be expressed in terms of the renormalization factor $Z$ and the effective range as  
\be\label{Eq:k1Z}
k_1\approx -\frac{i \ Z}{r},
\ee
where in the regime studied here, $ |{r}| \ll |{a}|$, $Z\ll 1$. Moreover,  $r$ is negative since range corrections are neglected.
It follows from Eqs.~\eqref{Eq:poles} and \eqref{Eq:k1Z} that, for a given  value of $r$,  the effect on the pole from changing  the scattering length  when some QCD parameter is varied is equivalent to that from changing  $Z$.
In case of $S$-wave scattering  there is a very close connection between shallow bound states and virtual states: as long as some minor change in the parameters makes an attractive  potential  weaker,  the corresponding  bound-state pole moves along the imaginary $k$ axis towards the threshold to turn eventually into a near-threshold pole  on the negative imaginary $k$ axis, that is to a virtual state. 
As follows from the pole $k_1$ in Eq.~\eqref{Eq:poles}, this transition corresponds to the change in the scattering length from $-\infty$ for a very shallow bound state to  $+\infty$ for a very shallow virtual state. 
At the same time,  we get from Eq.~\eqref{Eq:k1Z} that switching from a bound state to a virtual state calls for $Z$ to change
its  sign. The logic in this paragraph can actually be reversed. 
Let us assume that in some calculation  a pole moves smoothly from a bound to a virtual state leaving $r$ nearly constant when some QCD parameter is varied. 
Since this transition requires a sign change in $k_1$ and
since $Z$ is expected to change smoothly,  it follows from Eq.~\eqref{Eq:k1Z} that $Z$ needs to change from
small and positive to small and negative. 
Since $Z\approx 0$ points at a molecular state, the whole pole trajectory would correspond to a molecular or two-hadron scenario.  
Therefore, near-threshold virtual states necessarily have a dominating
two-hadron component and should be labeled as molecular as well.  The compactness $\bar Z_A$ (or compositeness $\bar X_A$) from Eq.~\eqref{eq: Zbar via a,r; own} is fully consistent 
with this conclusion  since  it treats  virtual states on an equal footing with bound states, that is $\bar Z_A \approx 0$ for $| {r} |\ll |{a}|$.

The other extreme case,  where  $| {a}| \ll |{r}|$ holds, corresponds to  compact states with $Z$ close to 1.  To have the poles near the origin in this case,  the small and negative scattering length 
must compensate for the large and negative effective range such that  $ar\approx 1/(\mu E_B)$. Then,  
one gets two  poles on the imaginary  axis nearly equidistant from the origin that employing the effective range 
expansion can be expressed as
\be\label{Eq:poles_compact}
k_{1,2}= \pm i \sqrt{\frac{2}{a\, r}}+ \frac{i}{ r} + {\cal{O}}\left(\sqrt{\frac{a}{r^3}}\right).
\ee
Alternatively, one can again express the poles in terms of  the effective range and $Z$ to get
 \be\label{Eq:poles_compactZ}
k_{1,2}\approx \pm \frac{i}{ r (1-Z)}+ \frac{i}{ r} ,
\ee
where, as before, the correction terms were dropped.
As follows from Eq.~\eqref{Eq:poles_compact}, when the scattering length changes sign the system  goes directly from
the bound state scenario to a resonance with poles located at (note: we then have $a>0$ and $r<0$)
\begin{equation}
  k_{1,2}^{\rm res.} \simeq \pm \sqrt{-\frac{2}{a\, r}}+ \frac{i}{ r}
  \label{eq:resonance}
\end{equation}
without going through a virtual state.  
Formally, this transition would require $Z$ to change from a real value close to 1 to a complex quantity  $Z= 1\pm i \sqrt{|a/2r|}$, while $\bar Z_A$ just gives $\bar Z_A =1- \sqrt{|a/2r|}\approx 1$  for $| {a}| \ll |{r}|$.
As long as the poles discussed above are located in the vicinity of  the threshold, 
the range corrections $\sim  \mathcal{O}(1/\beta)$ are not expected to change the conclusions qualitatively. 

We therefore found that, as long as range corrections can be neglected,
very narrow resonances very near, but above, an  $S$-wave threshold will generate poles as in Eq.~\eqref{Eq:poles_compact} and thus they should 
be predominantly of compact nature.
In particular, a compact resonance close to a continuum threshold should couple to this threshold
only very weakly. Denoting the poles, which are defined in the complex $E$ plane  relative to the threshold,  as $E^{\rm res.}$, from Eq.~\eqref{eq:resonance}, one has 
\begin{equation}
  \left|\frac{{\rm Re}\, E^{\rm res.}}{ {\rm Im}\, E^{\rm res.}}\right| \simeq \sqrt{\left|\frac{r}{2a} \right| } \gg 1.
  \label{eq:narrow}
\end{equation}
Thus,   resonances satisfying Eq~\eqref{eq:narrow} qualify as very narrow here.

Before closing this section we would like to confront the quantities $\bar Z_H$ and $\bar Z_A$ introduced in Eqs.~(\ref{eq:Hyodo}) and (\ref{eq: Zbar via a,r; own}), respectively, with the observations reported above. 
As was already discussed in Sec.~\ref{Sec:extension}, $\bar Z_H$ is close to zero in the regime where $|r| \approx a$, with $r<0$ and $a>0$, see Fig~\ref{fig:ZHvsZA}. On the contrary, from the above considerations it follows that, as long as range corrections are neglected, $ |r|\approx a$ is some interim regime, where both compact and two-hadron components are present in the state --- in line with $\bar Z_A\approx0.4$.

To summarize,  the condition, for a near-threshold state to contain a dominant two-hadron component, can be derived from the effective range parameter directly: if   $|r|\sim 1/\beta$, the state is predominantly molecular. If, on the other hand, a state is predominantly compact, then $r$ is negative and $|r| \gg 1/\beta$. 
These conditions hold for near-threshold bound states, virtual states as well as resonances and are encoded in the quantity $\bar Z_A$ introduced in Eq.~(\ref{eq: Zbar via a,r; own}). The information on the effective range can be inferred from studying the pole trajectories using  lattice QCD with the help of the appropriate effective field theories.

\section{Applications}
\label{Sec:appl}

\subsection{\texorpdfstring{$NN$}{NN} scattering} 

It was the deuteron for which Weinberg successfully applied his criterion for compositeness of near-threshold $S$-wave bound states~\cite{Weinberg:1965zz}. 
The deuteron is an isoscalar shallow bound state in the coupled $^3S_1$-$^3D_1$-channel. As expected, the deuteron turned out to be primarily molecular in nature, see Sec.~\ref{Sec:extension} for further discussion of this topic. There is also a pole in the isovector $^1S_0$-channel, but this pole is a virtual state.
In this channel, the effective range parameters read~\cite{Reinert:2017usi} 
\begin{align}
\label{Eq:param1s0}
	a &= 23.7 \ \text{fm} = 0.121 \ \text{MeV}^{-1},
\nonumber \\
	r &= 2.7 \ \text{fm} = 0.0136 \ \text{MeV}^{-1} \: .
\end{align}
They can be compared to the inverse of the typical momentum scale of the binding interaction, 
\begin{equation}
1/\beta  \sim  1/m_\pi  \approx   0.007 \ \text{MeV}^{-1} \ .
\end{equation}  
Since the effective range is positive and of the order of the range of forces, it is clearly a range correction. A possible short-range compact component that would provide a negative contribution to the effective range, is apparently smaller than the range corrections. Thus like the deuteron this virtual state is primarily composite. In line with this we find  $\bar X_A =0.9$.

In Ref.~\cite{Baru:2016evv},  an effective field theory  approach based on low-energy theorems in $NN$ scattering was employed  to analyze lattice results  at an unphysical pion mass $m_{\pi} =450$~MeV obtained by the NPLQCD collaboration~\cite{Orginos:2015aya}. 
At this pion mass both channels have bound states. In particular,  using the binding  energies of the deuteron and the dineutron system from Ref.~\cite{Orginos:2015aya},  the effective range parameters were extracted at next-to-leading order (NLO) in this EFT,  namely,
\begin{align}
\label{ERENLOfm}
a_{\rm NLO}^{(^3 \hskip -0.025in S _1)}   &=
-2.234({}^{+0.144}_{-0.156} ) \big({}^{+0.072}_{-0.052} \big) 
\text{ fm}, \nonumber \\
r_{\rm   NLO}^{(^3 \hskip -0.025in S _1)} \ &=\
1.07 \big({}^{+0.03}_{-0.03} \big) \big({}^{+0.08}_{-0.05} \big) \text{ fm}, \\
    a_\text{NLO}^{(^1\!S_0)}  &= 
-2.501\big({}^{+0.174}_{-0.481} \big) \big({}^{+0.304}_{-0.123} \big) 
\text{ fm},\nonumber \\
r_\text{ NLO}^{(^1\!S_0)} \ &=\ 
1.25 \big({}^{+0.05}_{-0.12} \big) \big({}^{+0.32}_{-0.12} \big) \text{ fm},
\end{align}
where  the uncertainties in the first and second brackets are statistical and systematic, respectively.
Using these parameters one finds that $\bar X_A \approx 0.7$ both in the  $^3S_1$ and $^1S_0$ channels, which is also consistent with a predominantly molecular scenario. Since the effective range is small ($r\sim 1/m_\pi$) and positive, also in this case, the deviation of $\bar X_A$ from unity is expected to be mostly from the range corrections.

\subsection{\texorpdfstring{$DK$}{DK} System}

In Ref.~\cite{Hanhart:2014ssa}, the pole trajectories for the $f_0(500)$, also known as $\sigma$, were studied as a function of the light quark (up and down) mass (for more on this see Ref.~\cite{Hanhart:2008mx,Albaladejo:2012te,Pelaez:2010fj}). 
There it was shown that for a sufficiently large quark mass the $\sigma$ meson becomes a bound state. The corresponding pole trajectories are fully in line with what was discussed above for $|a| \gg |r|\sim \beta$. Accordingly, at least for quark masses where the pole is close to the threshold, the $f_0(500)$ should be viewed as a hadronic molecule.

\begin{figure*}
\centering
\includegraphics[width=.45\textwidth]{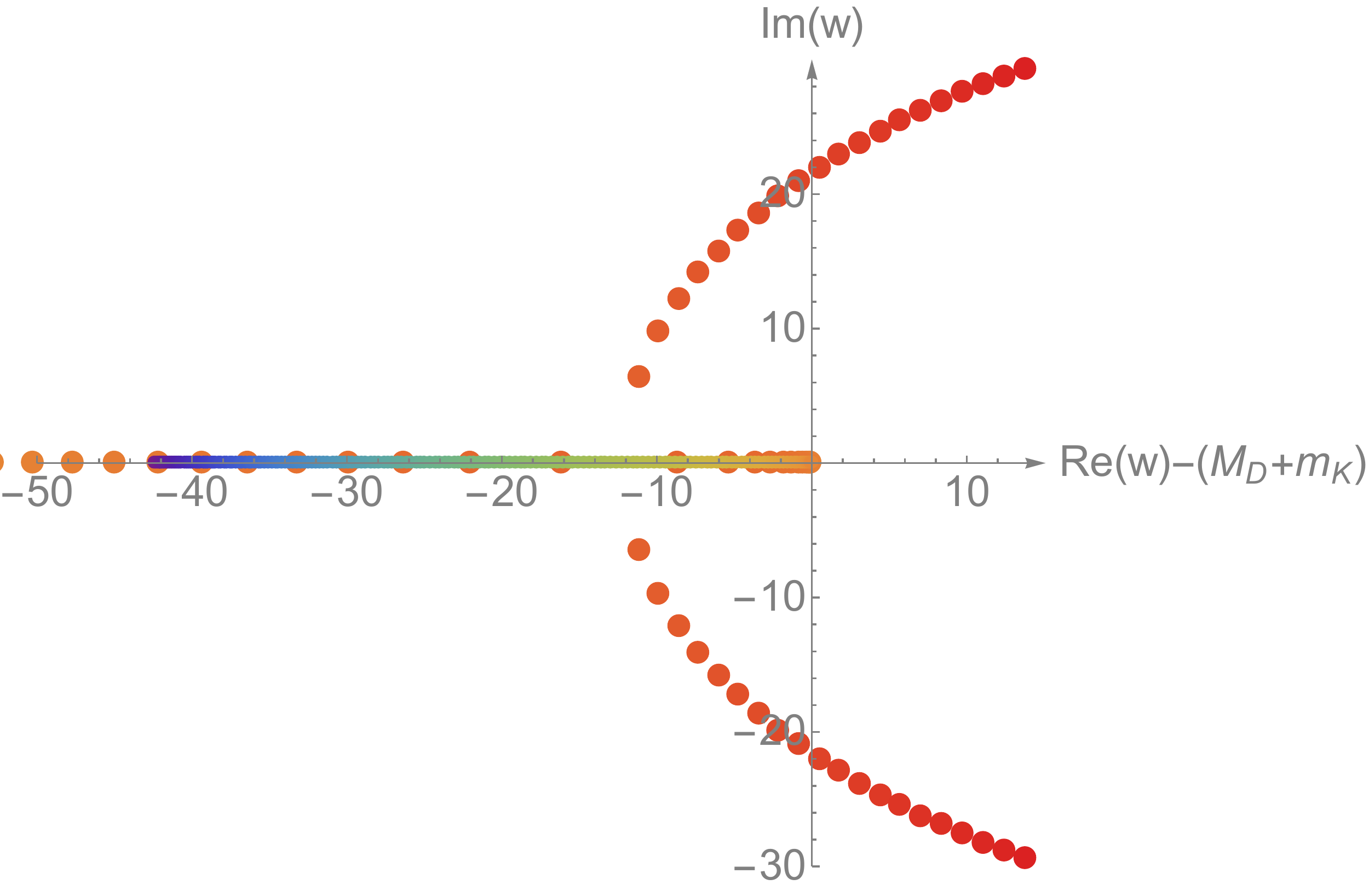}\hspace{0.3cm}
\includegraphics[width=.4\textwidth]{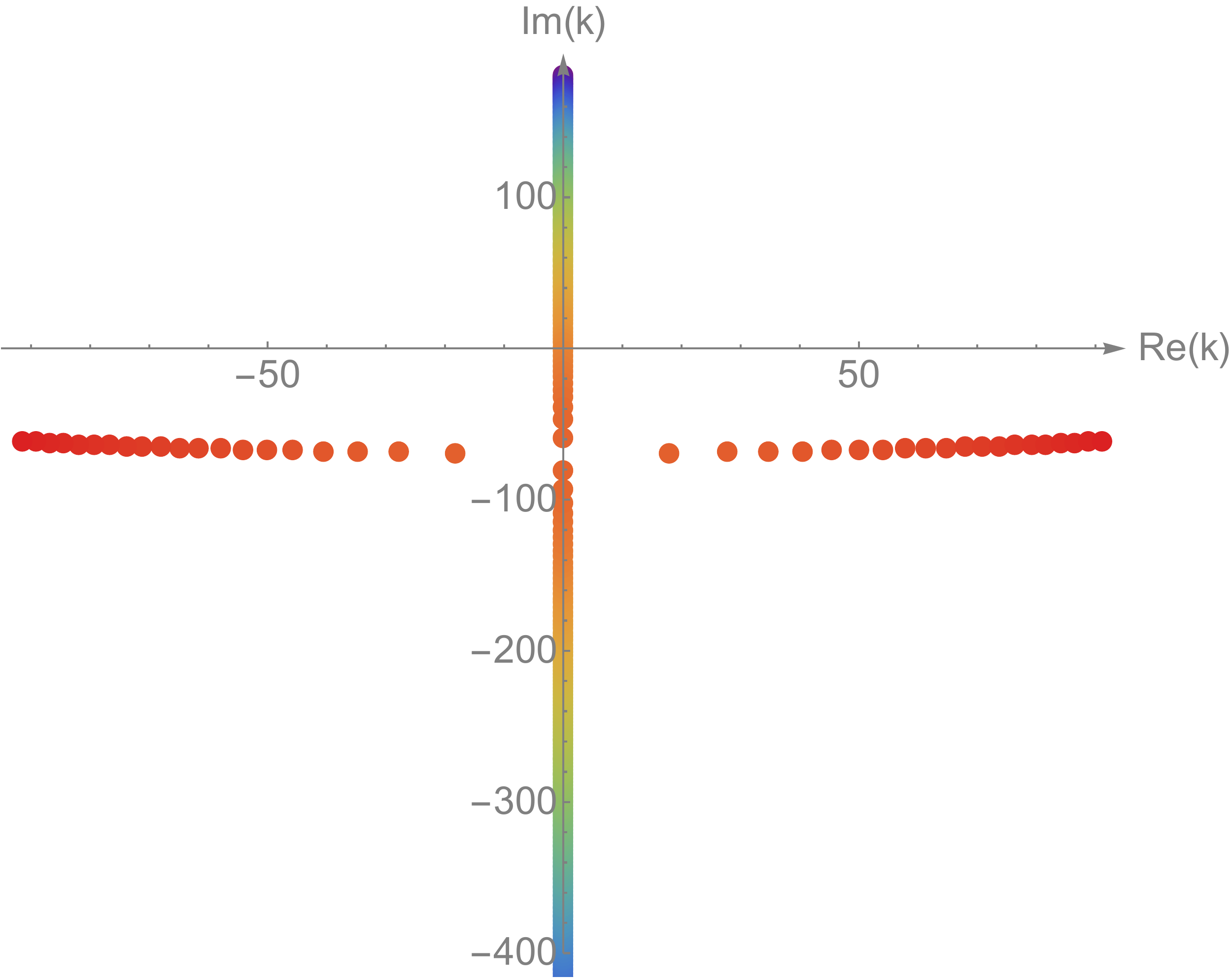}\hspace{0.3cm}
\includegraphics[width=.06\textwidth]{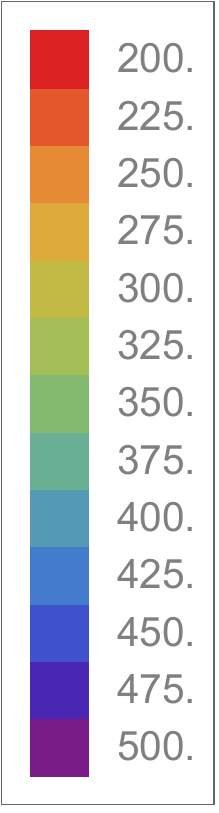}\\
\caption{Pole trajectories of the $\Ds$ when  the kaon mass is varied between $200$ MeV and $500$ MeV, as indicated by color:  a) Pole trajectories in the energy plane relative to the $KD$ threshold; b) Trajectories in the momentum plane.  Both the leading and more distant poles are shown in the plot. }
\label{fig:PoleTrajectories}
\end{figure*}

{\new An additional interesting case is the $DK$ system and its connection to the $\Ds$ pole. Since its discovery in 2003 by the BaBar Collaboration~\cite{Aubert:2003fg}, the positive-parity scalar charm-strange meson $\Ds$ with a very narrow width has been studied extensively. The mass of $\Ds$ lies only 45~MeV below the $DK$ threshold, which makes it a natural candidate for a hadronic molecule. 
That this interpretation also fits to recent $B$-decay data is discussed in Ref.~\cite{Albaladejo:2016hae}.
Additional arguments in favour of a molecular nature of the $\Ds$ as well as earlier references can be found in Refs.~\cite{Guo:2017jvc,Guo:2019dpg}.
}

To investigate the pole trajectories of the mentioned state, we utilize the analysis done in Ref.~\cite{Liu:2012zya} since it provides us with direct access to the quark mass dependencies of the system. In this work,  the L\"uscher finite volume technique is used to calculate the Goldstone-boson--$D$-meson scattering lengths in those channels that do not have disconnected diagrams. 
The resulting pion mass dependence of the scattering lengths is used to fix all low energy constants (LECs) that appear at NLO of the chiral expansion. The amplitude completely determined in this way was then used to calculate other channels,  in particular the isospin-0, strangeness-1 channel in which the $\Ds$ resides.
Here we investigate the dependence of the pole positions on the kaon mass. For larger-than-physical kaon masses this was discussed in Ref.~\cite{Cleven:2010aw}. 
Here we want to continue the kaon mass to smaller than physical values to observe the transition of the bound state to a virtual state. 

To find the kaon mass dependence of the involved states, let us consider the effective chiral Lagrangian for $D$-mesons and Goldstone bosons at NLO taking into account only the strong interaction. 
The LO Lagrangian is just the chirally covariant kinetic energy term of the heavy mesons
\begin{equation}
	\mathcal{L}^{(1)} = \mathcal{D}^\mu D^\dag \mathcal{D}_\mu D - m_D^2 D^\dag D \: ,
\label{eq: Lagrangian LO}
\end{equation}
with $D = (D_0,D^+,D^+_s)$ denoting the $D$-mesons, $m_D$ the $D$-meson mass in the chiral limit, and the covariant derivative being
\begin{align}
\mathcal{D}_\mu &= \partial_\mu + \Gamma_\mu \: , \nonumber \\
\Gamma_\mu &= \frac{1}{2} \left( u^\dagger \partial_\mu u + u \: \partial_\mu u^\dagger \right), 
\end{align}
where
\begin{equation}
U = \exp \left( \frac{\sqrt{2} i \Phi}{F_\pi} \right) \:, \quad u^2 = U \: .
\end{equation}
The Goldstone boson fields are collected in the matrix
\begin{equation*}
	\Phi = 
		\left( \begin{matrix}
			\pi^0/\sqrt{2} +  \eta/\sqrt{6} & \pi^+ & K^+ \\
			\pi^- & -\pi^0/\sqrt{2} +\eta/\sqrt{6} & K^0 \\
			K^- & \bar{K}^0 & -2\eta/\sqrt{6}
		\end{matrix}\right) \: .
\end{equation*}
Counting the $D$-meson masses as order $\mathcal{O}(p^0)$, the LO terms in the chiral Lagrangian are of $\mathcal{O}(p)$, and the NLO terms are of $\mathcal{O}(p^2)$.
The NLO chiral Lagrangian describing the interactions of the pseudoscalar charm mesons with the Goldstone bosons is given by
\begin{align}
	\mathcal{L}^{(2)} =& D \left( -h_0 \left< \chi_+ \right> - h_1 {\chi}_+ -h_2 \left< u_\mu u^\mu \right> - h_3 u_\mu u^\mu \right) \bar{D} \nonumber \\
	& + \mathcal{D}_\mu D \left( h_4 \left< u^\mu u^\nu \right> - h_5 \left\{ u^\mu u^\nu \right\}\right) \mathcal{D}_\nu \bar{D},
\label{eq: Lagrangian NLO}
\end{align}
where
$	\chi_+ = u^\dagger \chi u^\dagger + u \chi u \  \mbox{and} \
	u_\mu = i u^\dagger \partial_\mu U u^\dagger \: .
$
The quark mass matrix enters via
\begin{equation}
 \chi = 2 B \cdot \text{diag}\left( m_u, \: m_d, \: m_s \right) \ ,
\end{equation}
where $B = \left| \left< 0 | \bar{q} q | 0 \right> \right| / F_\pi^2$. Further, $F_\pi$ is the pion decay constant in the chiral limit.
Since the interaction is quite strong in some channels, and in one channel even a bound state is produced, the interactions derived from the Lagrangian given above are resummed using the Lippmann-Schwinger equation.
The LEC $h_1$ can be fixed from the SU(3) splitting between the $D_s^+$ and $D^+$ masses, and the LECs $h_i$ with $i = 0, 2,\ldots, 5$, can be fixed by a fit to lattice data~\cite{Liu:2012zya}.
The resulting amplitudes, that contain poles in various channels, turn out to be also consistent with the momentum dependence of the $I=1/2$, non-strange 
scattering amplitudes~\cite{Albaladejo:2016lbb} determined recently in lattice QCD~\cite{Moir:2016srx}, and can describe the LHCb measured $D\pi/D\bar K$ distributions for a series of three-body $B$ meson decays~\cite{Du:2017zvv,Du:2019oki}.
Since in the Lagrangians above all light quark mass dependence is explicit,
we are now in the position to investigate systematically the quark mass
dependence of the system.
Varying the up and down quark masses (the strange quark mass) is equivalent to varying 
the pion mass (kaon mass).  As we are interested in the transition of the $\Ds$ pole to a virtual state, we lower the kaon mass.

Using the Lagrangian given in Eq. (\ref{eq: Lagrangian NLO})
we find the NLO correction to the charmed meson masses to be 
\begin{align}
  M_D &= {M}_D^{\rm phys.} + 2 h_0\frac{M_K^2-M_K^{\rm phys. \, 2}}{{M}_D^{\rm phys.}},   \\
 M_{D_s} &= {M}_{D_s}^{\rm phys.} + 2 (h_0 + h_1) \frac{M_K^2-M_K^{\rm phys. \, 2}}{{M}_{D_s}^{\rm phys.}}, 
\end{align}
where we considered isospin symmetry and used
$M_K^2 = B (m_s + \hat m)$
for the kaon mass with $\hat m = (m_u+m_d)/2$. The values of $h_0$ and $h_1$ were determined to be $h_0=0.014$ and $h_1=0.42$~\cite{Liu:2012zya}.
Finally, following reference \cite{Cleven:2010aw} we use that the eta mass is given to this order by the Gell-Mann--Okubo relation, 
\begin{equation}
M_\eta^2 = \frac{4}{3} M_K^2 - \frac{1}{3} M_\pi^2 \: .
\label{eq:GellMannOkubo}
\end{equation}

Note that in the fits of Ref.~\cite{Liu:2012zya} the physical value for the pion decay constant, $F_\pi = 92.21$ MeV, was used,
since the difference from its chiral limit value, and hence its pion or kaon mass dependence, is a higher-order effect. 
We focus on the sector that contains the $\Ds$ and therefore need to work with two coupled channels, namely $DK$ and $D_s \eta$.
The  resulting pole trajectories are shown in  Fig.~\ref{fig:PoleTrajectories}. 
At the physical kaon mass ($m_K\approx 500$ MeV), the system is characterized by a 
very asymmetric pair of poles, one on the first sheet close to the threshold and the other on the second sheet far away from the threshold (too distant to be seen in the plot). Thus, the $\Ds$ should indeed be interpreted as a hadronic molecule.
As the kaon mass is lowered, the two poles start to approach each other. At a kaon mass of about 250 MeV the leading pole moves from the first to the second sheet: at this mass the $\Ds$ turns into a virtual state. 

Thus, as discussed in Sec.~\ref{Sec:poles}, the bound state does indeed approach the threshold and then switches over to the second sheet to form a 
virtual state when the kaon mass is lowered. At the physical kaon mass, the effective range parameters are given by $a = -0.86$~fm and $r =- 0.45$~fm which gives $Z= \bar Z_A=0.30$ or $X= \bar X_A=0.70$, employing
either of  Eqs.~(\ref{eq: Z from ERE}) or \eqref{eq: Zbar via a,r; own}.
Moreover, as will be discussed at the end of this section, at the physical kaon mass for the system at hand, $\gamma=134$~MeV and $\beta\sim |1/r^{\rm eff.}|=330$~MeV, where $r^{\rm eff.}$ is defined in Eq.~\eqref{reffdef} below, induced by the nearest threshold, such that $(\gamma/\beta)=0.4$, which quantifies the uncertainty.  
Thus, according to the Weinberg criterion, the $D_{s0}^* (2317)$ is even compatible with purely molecular state. When we lower the kaon mass, the compactness $Z$ lowers further until it reaches $0$, when the pole hits the threshold.
After this point, formally the Weinberg criterion can no longer be applied as discussed above. 
However, the compactness is expected to be continuous under changes in a fundamental parameter as the quark masses. Thus, just after the transition to the second sheet, the state is still mostly molecular in nature, which is captured in the extended compactness $Z_A$.

\begin{figure}
\centering
\includegraphics[width=\linewidth]{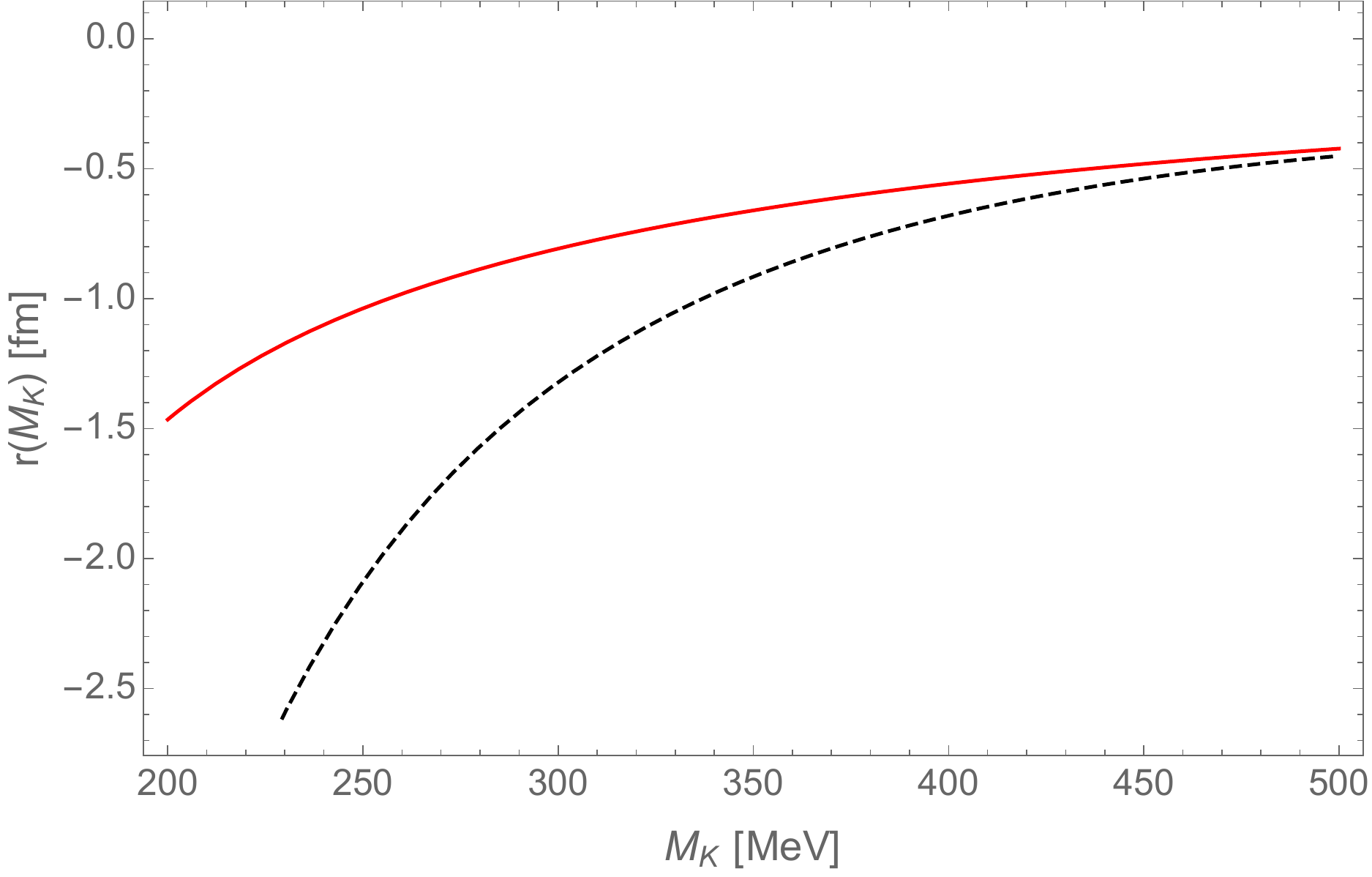}\\
\caption{Comparison of the effective range extracted from the scattering amplitudes
(black dashed line) with $r^{\rm eff.}$ (red solid line). }
\label{rcomparison}
\end{figure}

For the system at hand, it is interesting to have a closer look at the origin of the negative effective range reported above.
Because up to and including NLO there are only point interactions, 
there is no obvious range in the system. From this, one could expect $r\approx 0$ in the molecular case,
and, accordingly, some negative effective range might be interpreted as some admixture of a
compact component. 
However, an additional scale enters through the $D_s \eta$ channel.  In the effective range expansion of the $DK$ scattering amplitude, this channel
enters predominantly via the analytic continuation of the unitarity cut contribution, namely
\begin{align}
\nonumber
&-i\frac{g_{D_s\eta}^2}{g_{DK}^2}\sqrt{2\mu_{D_s\eta}\left(M_D{+}M_K{+}\frac{k^2}{2\mu_{DK}}{-}\left(M_{D_s}{+}M_\eta\right)\right)} \\ \nonumber
& \quad\quad =\frac{g_{D_s\eta}^2}{g_{DK}^2}\sqrt{2\mu_{D_s\eta}\left(\Delta M{-}\frac{k^2}{2\mu_{DK}}\right)} \\
& \quad\quad =: \frac{g_{D_s\eta}^2}{g_{DK}^2}\sqrt{2\mu_{D_s\eta}\ \Delta M}+\frac12 r^{\rm eff.} k^2 + {\cal O}(k^4) \ ,
\label{otherthreshold}
\end{align}
where $\Delta M=M_{D_s}{+}M_\eta{-}M_D{-}M_K$ and $g_{D_s\eta}$ and $g_{DK}$ denote
 the
coupling of the given partial wave to the $D_s\eta$ and $DK$ channel, respectively,  and we defined
\begin{equation}
r^{\rm eff.}=-\frac{g_{D_s\eta}^2}{g_{DK}^2}\sqrt{\frac{\mu_{D_s\eta}}{2\mu_{DK}^2\Delta M}} \ .
\label{reffdef}
\end{equation}
Thus, the sign of the term in Eq.~(\ref{otherthreshold}) is fixed by unitarity. 
To estimate the ratio of the couplings, we may use the strengths of the $DK\to DK$ and $DK\to D_s\eta$
potentials at LO. With these the ratio of the squared couplings in Eq.~(\ref{reffdef}) is 3/4 (see, e.g., Ref.~\cite{Guo:2006fu}).
In Fig.~\ref{rcomparison}, the effective range extracted from the $DK$ scattering amplitude directly is
compared to the effective parameter defined in Eq.~(\ref{reffdef}). Clearly, the effective range
parameter is largely provided by the $D_s\eta$ channel.
For small values of $M_K$ one finds that $|r^{\rm eff.}|$ gets large, since $\Delta M$ gets very small,\footnote{The $D_s\eta$ and $DK$ thresholds collide at $M_K\simeq130$~MeV; decreasing the kaon mass further, the $D_s\eta$ threshold would be the lower one.}
which at the same time leads to a bad convergence of the ERE for the single-channel $DK$ scattering amplitude.
Clearly, as soon as the two thresholds get close together, an estimate given in Eqs.~\eqref{otherthreshold} and \eqref{reffdef} that treats dynamically only a single two-hadron
channel is no longer applicable to interpret the findings from the coupled-channel calculations discussed above.
In such a case, a coupled channel generalisation of the Weinberg formalism is needed to study the molecular
admixture in a given state, as discussed, e.g., in Refs.~\cite{Gamermann:2009uq,Hyodo:2011qc,Aceti:2014ala,Guo:2015daa}.  
The discussion above shows how the onset of a second channel becomes visible in an analysis of the Weinberg type.

\section{Further Discussions and Disclaimer}
\label{disclaimer}

It was mentioned in various places that range corrections can in some cases modify
significantly the results, especially for molecular states. This was already discussed above
in case of the $\Ds$, where the range corrections to the $Z$ parameter were estimated to
be as large as 40\%.
Recent studies showed~\cite{Wang:2018jlv,Baru:2019xnh} that
range corrections can play an even more important role in  composite systems made of two heavy mesons,
 as soon as the one-pion exchange is allowed to contribute. The reason for this is two-fold:
On the one hand, these doubly-heavy systems are often driven by coupled channels
(e.g. $B\bar B^*$ and  $B^*\bar B^*$ for $Z_b(10610)$ and $Z_b(10650)$),
such that the neighbouring channel can induce large relative momenta, which would  imply large contributions to the effective range in line with Eq.~\eqref{otherthreshold}.
On the other hand, because (a) the one-pion exchange comes with a tensor force and (b)  the coupled-channel momenta are large, 
coupled-channel transitions to $D$-waves are strong and, because of the small mass of the pion, 
not  kinematically suppressed. This mechanism can even push the poles above the threshold.
Nevertheless, although the imaginary parts of the resonance poles in the complex $E$ plane obtained in Ref.~\cite{Baru:2019xnh} are all less than about 20~MeV, they satisfy $|{\rm Re}\, E^{\rm res.}/{\rm Im}\, E^{\rm res.}| \lesssim 1$. In view of Eq.~\eqref{eq:narrow}, they still qualify as molecular states.

{\new The derivation of the Weinberg compositeness criterion and its extensions generally relies on that the scattering amplitude near the threshold fulfils the effective-range expansion,  
which is expected to be valid for the momenta much smaller than the inverse range of forces. 
While in most cases this is indeed correct, under certain specific conditions, the interplay between quark states and a nonperturbatively interacting hadron–hadron continuum 
may cause  this picture to fail.  In particular, in Ref.~\cite{Baru:2010ww}   it was shown that  the scattering amplitude may also have zeros in the near-threshold region if the following criteria are met:  
 a) the hadronic  interaction is sufficiently strong to support a bound or virtual state  
 and b) a   quark state exists near the threshold with a weak coupling to the hadronic channel.  The appearance of such a zero invalidates the effective-range expansion and corresponds to three 
 near-threshold poles in the scattering amplitude.  Such a scenario goes beyond the scope of the present study.  The reason is that with the pole locations of several states near the threshold there 
 is more than one small scale in the system which may circumvent a controlled expansion. 
An example, where such a case could occur was considered in Ref.~\cite{Cincioglu:2016fkm}, where the interplay of the $X(3872)$ and the  charmonium state $\chi_{c1}(2P)$, controlled by some mixing parameter, was 
considered, and accordingly several poles were found.   
However, even the results of Ref.~\cite{Cincioglu:2016fkm} for the compositeness of the $X(3872)$ are fully in line 
with the present analysis at least for extreme values of the mixing parameter.   
}

 {\new In many theoretical analyses of data with near threshold structures, 
an effective Lagrangian approach is employed, 
which does not involve explicit bare poles.
It is assumed that if additional poles exist, they are not in the vicinity of thresholds and thus can be accounted for perturbatively via a tower of (energy-dependent) contact interactions.
The near-threshold state in such a dynamical approach can be interpreted as molecular or compact even without the explicit inclusion of a quarkonium component, see also Ref.~\cite{Baru:2019xnh} for a related discussion. 
 In this context, let us mention an application of Eq.~\eqref{eq: Zbar via a,r; own} to a new state predicted near the  $J/\psi J/\psi$ threshold in Ref.~\cite{Dong:2020nwy} from an analysis of the recent LHCb data on the double-$J/\psi$ spectrum~\cite{Aaij:2020fnh}.
In Ref.~\cite{Dong:2020nwy}, it was argued that the data can be  well described using two coupled-channel potential models 
(employed for  constructing the $T$-matrix via the Lippmann--Schwinger equation): a two-channel ($J/\psi J/\psi$ and $J/\psi \psi(2S)$) model with an energy-dependent potential, and a three-channel ($J/\psi J/\psi$, $J/\psi \psi(2S)$ and $J/\psi\psi(3770)$) model with a constant potential. In both models, a pole near the double-$J/\psi$ threshold was found. Though no pre-existing bare pole was introduced, 
whether this  state is molecular or compact   depends strongly on the dynamical mechanisms proposed: the use of energy-dependent interactions yields predominantly a compact state while a three-channel mechanism  suggests its molecular nature. 
}

\section{Conclusion}

In this paper, we use certain plausible smoothness assumptions to extend the well established
Weinberg criterion, introduced to characterize the compositeness of bound states, to virtual states
and resonances.  In particular, we demonstrate that an  extension, proposed in 
Ref.~\cite{Hyodo:2013iga}, is inconsistent with these assumptions, while a slightly modified
formula for the compositeness, namely $$ \bar{X}_A =  \sqrt{\frac{1}{1 + \left| {2r}/{a} \right| }},$$ is consistent with all limiting cases and provides a smooth 
transition from the regime of molecules to the regime of compact states.
By analyzing the pole trajectories and their relation with the compactness, we find 
that near-threshold virtual states are of molecular nature while narrow, near-threshold resonances should be interpreted as predominantly compact, as long as range corrections can be neglected. 
Further, we argue that the  compositeness originally proposed by Weinberg as well as  some of  its extensions   contain a pole at $r=-a/2$. As a consequence, this criterion cannot be applied 
to systems having  bound states with  positive effective ranges  
even for a natural case when the range corrections are suppressed relative to the binding momentum. This is illustrated using the deuteron as an example. 
On the other hand, the compositeness proposed in this study  is by construction free of this pole and can be used  to obtain reasonable estimates up to the range corrections for the compositeness of bound states, virtual states and resonances. 
We consider several applications. First, we focus on $NN$ scattering, where the range corrections are of the order of the inversed pion mass, to demonstrate that 
this system is predominantly composite both at physical and unphysical pion masses.  
In addition, we discuss the $DK$ isoscalar system with the emphasis on the pole trajectory of the $\Ds$ state.
This system is of particular interest here, since the leading range corrections do not come from
the exchange of a meson in the $t$-channel, but from the nearest  $D_s \eta$ channel residing above the $DK$ threshold and coupled strongly
to this system. It is demonstrated that the contribution to the effective range  in such a case  
is always negative. Thus, if the Weinberg criterion is applied naively to such a system the effect
of the other threshold can mimic a compact component although, as argued in this study, it should be interpreted as a range correction.

Finally,   we also argue that in some special cases (especially for molecular states)
the values for the compositeness can be changed significantly by range corrections.
This is in particular true for the  systems involving several particle coupled channels, where the pion $t$-channel exchange including its sizeable 
$S-D$ transitions driven by the tensor force contributes. In those situations the conclusions
described above need to be modified and adapted. A detailed study of this case requires further research
that goes beyond the scope of this work.

\medskip

\begin{acknowledgements}

This work is supported in part by the National Natural Science Foundation of China (NSFC) and  the Deutsche Forschungsgemeinschaft (DFG) through the funds provided to the Sino--German Collaborative Research Center  CRC110 ``Symmetries and the Emergence of Structure in QCD" (NSFC Grant No.~11621131001), by the NSFC under Grants No.~11835015, No.~11961141012, and No. 11947302, by the Chinese Academy of Sciences (CAS) under Grants No. XDB34030303 and No. QYZDB-SSW-SYS013, by the CAS Center for Excellence in Particle Physics (CCEPP)  and by the Russian Science Foundation (Grant No. 18-12-00226).

\end{acknowledgements}

\begin{appendix}

\section{Wigner bound}
\label{sec:app}

The Wigner's causality inequality reads (see, e.g., Ref.~\cite{Bohm})
\begin{equation}
	\frac{d\delta_l}{dk} \geq - R + (-1)^l \frac{\sin2(\delta_l+kR)}{2k},
	\label{eq:wigner}
\end{equation}
where $\delta_l$ is the phase shift for the $l$-th partial wave, and $R$ is the interaction radius. For the $S$-wave, considering the ERE 
\begin{equation}
	k\cot\delta_0 = \frac1{a} + \frac12 r k^2,
\end{equation}
we have
\begin{align}
	\frac{d\delta_l}{dk} = \frac{\sin(2\delta_0)}{2k}  - r\sin^2\delta_0.
\end{align}
From Eq.~\eqref{eq:wigner}, we obtain the following bound for the effective range
\begin{align}
  r &\leq \frac1{\sin^2\delta_0} \left[R- \frac1{k}\sin(kR) \cos(2\delta_0 + kR) \right] \nonumber\\
  &\leq \frac{2R}{\sin^2\delta_0},
\end{align}
noticing that $1 - \frac1{kR}\sin(kR) \cos(2\delta_0 + kR) \in [0,2]$.
One sees that for a zero-range interaction, $r$ must be negative semi-definite,  
and for a finite-range interaction $r$ is smaller than some positive number.

\end{appendix}

\end{document}